\documentclass[fleqn,10pt]{wlscirep}

\usepackage{amsmath,graphicx}
\usepackage{epstopdf}
\usepackage{overcite}
\usepackage{authblk}

\usepackage{ulem,cancel}
\newcommand{\bra}[1]{\left \langle #1 \right \vert}
\newcommand{\ket}[1]{\left \vert #1 \right \rangle}

\begin{document}
%%%%%%%%%%%%%%%%

\title{A Nonequilibrium Variational Polaron Theory to Study Quantum Heat Transport}

\author[1,2]{ChangYu Hsieh}
\author[2]{Junjie Liu}
\author[2]{Chenru Duan} 
\author[1,2]{Jianshu Cao*}
\affil[1]{Department of Chemistry, Massachusetts Institute of Technology, 77 Massachusetts Avenue, Cambridge, MA 02139, USA}
\affil[2]{Singapore-MIT Alliance for Research and Technology (SMART), 1 CREATE Way, Singapore 138602, Singapore}
\affil[*]{Corresponding author: jianshu@mit.edu}

%\maketitle    
       
\begin{abstract}
We propose a nonequilibrium variational polaron transformation, based on an ansatz for nonequilibrium steady state (NESS) with
an effective temperature,   to study quantum heat transport at the nanoscale.  
By combining the variational polaron transformed master equation with the
full counting statistics, we have extended the applicability of the polaron-based framework to study nonequilibrium
process beyond the super-Ohmic bath models. Previously, the polaron-based framework for quantum heat transport 
reduces exactly to the non-interacting blip approximation (NIBA) formalism for Ohmic bath models due to
the issue of the infrared divergence associated with the full polaron transformation. The nonequilibrium variational method
allows us to appropriately treat the infrared divergence in the low-frequency bath modes and explicitly include cross-bath correlation
effects. These improvements provide more accurate calculation of heat current than the NIBA formalism for Ohmic bath
models.  We illustrate the aforementioned improvements with the nonequilibrium spin-boson model in this work and
quantitatively demonstrate the cross-bath correlation, current turnover, and rectification effects in quantum heat transfer.

\end{abstract}
%The variational approach also extends the applicability of polaron-based method beyond the scaling limit. 

\maketitle  

%\flushbottom
 
%\newpage  
%\tableofcontents

\newpage
            
%%%%%%%%%%%%%%
\section{Introduction}\label{sec:intro}
%%%%%%%%%%%%%%
% Talk about why Heat Important in Nanoscale
As modern electrical, optical and mechanical devices\cite{Nitzan.06,Nitzan.03.Science}
continue to shrink in size, nanoscale heat transfer has become 
an increasingly important research direction.  A thorough understanding and characterization of heat dissipation and fluctuation will be critical to maintain the stability of nanoscale devices.  For instance, better design of nanostructures to avoid joule heating\cite{Lee.Nat.13} could be realized with theoretical insights gained from studies of heat transfer in model systems.  Furthermore, the advancement of novel technologies such as phononics\cite{Li.12.RMP} and quantum heat engines\cite{Kosloff.14.ARPC,Xu.16.NJP,Dorfman.18.PRE}
 have spurred further interests in acquiring a precise 
control\cite{Leitner.15.AP,Narayana.12.PRL,Georg.19.PRB} of heat flows at nanoscale.   Beyond the application driven needs, the quantum transport also provides an experimentally accessible platform to explore the rich set of nonequilibrium physical phenomena\cite{Esposito.09.RMP,Campisi.11.RMP,Liu.17.JCP} in the quantum regime.  For instance, the fluctuation theorem for charge transport have been experimentally 
verified in quantum dot systems
\cite{Fujisawa.06.S,Kung.12.PRX,Gustavsson.06.PRL}.

% Recent Polaron Transform advantages and disadvantages
Motivated by the aforementioned interests, our group has recently developed a novel approach,
based on a combination of the nonequilibrium polaron transformed Redfield equation\cite{Wang.15.SR,Wang.17.PRA} (NE-PTRE) and the full counting statistics, to study heat transfer and its higher order moments for a
finite-size quantum system simultaneously coupled to two heat reservoirs held at different temperatures.  
The NE-PTRE approach inherits the physically transparent structure of the Redfield equation that facilitates the analyses of the underlying heat-conduction physics.  For instance, the parity classified transfer processes\cite{Wang.15.SR} through a two-level spin junction has been unraveled within the polaron framework.  More importantly, the NE-PTRE addresses the shortcomings of the standard Redfield equation approach which perturbatively treats the system-reservoir interactions in the weak coupling limit. In Ref.~\citenum{Wang.15.SR}, it has been demonstrated that the NE-PTRE provides an analytical expression for heat current interpolating accurately from the weak to strong coupling regimes for a super-Ohmic bath model.  This successful unification of heat current calculations builds upon the physical picture that bath modes displace to new stable configurations under coupling to the system. 
However, the polaron technique\cite{Lee.12.JCP,Lee.15.JCP,Xu.16.FP} performs rather poorly with the slower bath modes as they are sluggish and fail to dress the system.
Hence, potential problems arise when the polaron technique is applied to Ohmic and sub-Ohmic bath models which feature more prominent contributions from the low-frequency modes (i.e., the infrared divergence).  Indeed, it can be easily shown that the NE-PTRE reduces to the Nonequilibrium Non-Interacting Blip Approximation\cite{Nicolin.11.JCP} (NE-NIBA) results and captures only the incoherent part of the heat transfer processes for Ohmic baths.

% The improvement obtained in this work and the Y operator based on variational
Inspired by earlier studies on open quantum systems, the variational polaron transformation\cite{Silbey.84.JCP,Harris.85.JCP} is adopted in this work to extend the applicability of the NE-PTRE beyond the super-Ohmic bath models.  Under the variational treatment, slow modes are displaced with reduced displacements.  Technically, this variational modulation avoids the infrared divergence of a full polaron transformation when applied to Ohmic baths. The optimized dispaclements are determined by minimizing the upper bound of an effective free energy based on the Feynman-Bogoliubov inequality.  This variational ansatz only holds strictly for the equilibrium systems.  In this work, we extend this equilibrium technique to the nonequilibrium domain.  In subsequent discussions, we  denote the improved method nonequilibrium variational polaron transformed Redfield equation (NE-VPTRE). 
As confirmed by the numerical study, the 
NE-VPTRE method provides significantly more accurate results than that of the original NE-PTRE.  
In particular, the polaron-based calculation of heat current is now pushed beyond the NIBA accuracy 
for Ohmic and sub-Hhmic bath models. This achievement enables a unified heat current calculation 
(from weak to strong couplings) for all spectral densities of the thermal bath.

%Finally, we stress that the non-equilibrium
%variational method proposed in this work can be easily generalized beyond the NESB model.

The paper is organized as follows.
In Sec.~II,	we discuss the formal structure of nonequilibrium steady state 
and define the effective temperature for quantum transport, and then we
 introduce the  non-equlibrium spin-boston (NESB) model	and outline the derivation of NE-VPTRE.
In Sec.~III,  we analyze the proposed nonequilibrium variational method,
bechmark the theoretical predictions with the numerical exact results for NESB models,
and discuss the considerable advantages of the variational polaron transformation
over the usual polaron transformation.  Finally,	we present a brief summary in Sec.~IV.

%:Figure 1
\begin{center}
\begin{figure}
\includegraphics[width=0.75\columnwidth]{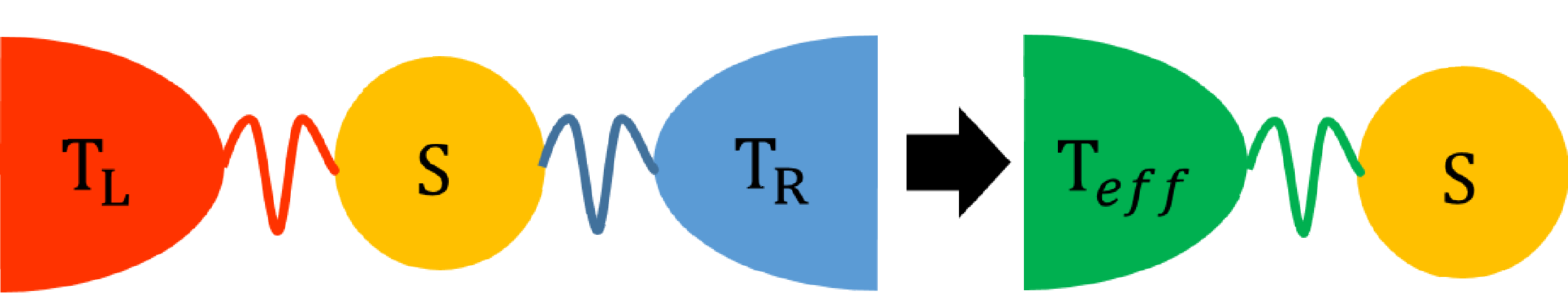}%width=0.9\columnwidth
		\caption{Schematics of a  quantum heat transfer model in which an open system 	(yellow circle, a two-level system for example) 
		is simultaneously coupled to two heat reservoirs
		held at two different temperature $T_L$ and $T_R$. 
	The temperature gradient induces heat current directed from hot to cold reservoir. 
	This non-equlibrium transport setup can be mapped to an effective equilibrum setup,
		so that the non-equlibrium steady state (NESS) is formally 
		written as a Gibbs state at an effective temperature. }
	\label{fig:model}
\end{figure}
\end{center}

%%

%%%%%%%%%%%%%%
\section{Method}
%%%%%%%%%%%%%%
%Nonequilibrium Steady States For Heat Transport}\label{sec:ness}
% Discuss the variational-adapted polaron with Full Counting
%%%%%%%%%%%%%%%%%%%%%%%%%%%%%%%%%%%%%%%%%%%%%%%%%%%%%%%%%%%%%
\subsection{Nonequilibrium Steady State}\label{sec:ness}
%%%%%%%%%%%%%%%%%%%%%%%%%%%%%%%%%%%%%%%%%%%%%%%%%%%%%%%%%%%%%
%%
%In this work, we set $\hbar=1$ and $k_B=1$. 

We illustrate the general non-equilibrium transport setup with the example of heat transfer,
	as illustrated in Fig.~\ref{fig:model}.
Here,   an open system couples directly to two thermodynamic reservoirs held at different temperatures,
and the long-time limit of the system defines the nonequilibrium steady state (NESS), which 
	is characterized by the heat flux from the hot to cold  baths.
This nonequilibrium phenomenon can be described by a general Hamiltonian,
\begin{align}
	H	&=	H_s + H^L_B +H^R_B+ H_I   ,
\end{align}
	where $H_s$ is the open system,
 $H^R_B$ and $H^L_B$  denote  the left and right bosonic reservoirs,
%are the  = \sum_{v=L,R} H^v_B = \sum_{k,v}\omega_{k,v} b^\dag_{k,v}b_{k,v}$
%	denotes the left and right bosonic reservoirs 
%	with $b^\dag_{k,v}$ and $b_{k,v}$ the creation 
%	and annihilation operators
%	for the $k$--th mode 
%	in the $v$--th bath.
and $H_I$ describes the system-reservoir interaction.
The two heat reservoirs are held fixed
	at the inverse temperature $\beta_R$ and $\beta_L$
	respectively.
This setup encompasses a broad range of dissipative
	and transport settings.
In this work, we set $\hbar =1$ and $k_B = 1$.

Our approach is built upon the theoretical concept\cite{Hershfield.93.PRL,NESS.14.PRE} that claims the formally exact nonequilibrium steady state (NESS), such as the heat transfer model with a temperature bias, can be cast in a Gibbs-like expression. 
This effective global equilibrium state 
(for the open system and its two heat baths) 
is augmented with an extra term, often referred to as the $Y$ operator, which corresponds to the entropy production associated with the irreversible  transport such as the heat transfer driven by a temperature gradient.  This formal structure reads
\begin{eqnarray}\label{eq:nessf}
 \rho_{NESS} 
&=& 
	\frac{e^{-\bar{\beta}\left(H+\frac{\Delta \beta}{\bar\beta} Y\right)}}
	%--------------------------------------------------------------------------
	{\text{Tr} e^{-\bar{\beta}\left(H+\frac{\Delta \beta}{\bar\beta}Y\right)} } ,
\end{eqnarray}
where $\bar\beta$ is an effective temperature and 
\begin{eqnarray}
Y = \frac{1}{2} \Omega \left(H_B^L  - H_B^R\right) \Omega^{-1},
\end{eqnarray}
$\Omega =\lim_{t\rightarrow \infty} \exp(itH)\exp(-itH_0)$ 
and 
$\delta\beta = \beta_L-\beta_R$.

Equation (\ref{eq:nessf}) is a joint density matrix for the open system and its environment. 
The close resemblance of Eq.~(\ref{eq:ness}) to
	a thermal equilibrium state implies that various nonequilibrium quantities can be computed by adopting the standard equilibrium techniques for observable calculations, 
	provided one has access to the exact form of the $Y$ operator.  
In general, 
	this is an extremely difficult task except for the simplest models.
In this work, 
	we simply exploitH the knowledge of the formal 
	structure\cite{Thingna.14.JCP,Dhar.12.pre,NESS.14.PRE} 
	of NESS and devise a variational polaron transformation technique 
	for quantum transport problems. 

\subsection{Mapping of Non-equilibrium Spin-Boson Model}

For convenience, we first introduce the non-equlibrium spin-boson (NESB) model, which is used throughout the paper
to formulate and calibrate the NE-VPTRE approach. 
The model Hamiltonian\cite{Leggett.87.RMP,Weiss.12.NULL,Thoss.14.JPCA,Nicolin.11.JCP,Kilgour.19.JCP} is given by
\begin{eqnarray}\label{eq:H_nesb}
H	& = & H_0 + H_I \nonumber \\
 	& = & H_s + H_B + H_I \nonumber \\
	& = & \frac{\Delta}{2} \sigma^x 
		+ \frac{\epsilon}{2} \sigma^z 
		+ H_B + \sum_v \sigma^z \otimes B_v.
\end{eqnarray}
On the first line, the Hamiltonians $H_0=H_s+H_B$ 
	and $H_I$ describe the uncoupled system $(s)$ and bath $(B)$
	and their coupling, respectively.
On the third line, we specify the system as a two-level spin with $\sigma^{z/x}$ referring to the standard Pauli matrices. 
$H_B=\sum_{v=L,R} H_B^v=\sum_{k,v}\omega_{k,v}b^{\dagger}_{k,v}b_{k,v}$ 
denotes the left and right bosonic reservoirs with $b_{k,v}^{\dagger}$ 
and $b_{k,v}$ the creation and annihilation operators for the $k$-th mode in the $v$-th bath. 
The two heat reservoirs are held fixed at the inverse temperature $\beta_v$, respectively.  
For the NESB model, the reservoirs are linearly coupled to the system as 
	$B_v = \sum_{k} g_{k,v} (b_{k,v}^\dag + b_{k,v})$. 
The model is summarized in Fig.~\ref{fig:model}.  
Though formulated in the context of the NESB model, 
we shall emphasize that the subsequent derivations of the effective temperature in Eq.~(\ref{eq:ness}) and NE-VPTRE are completely general and are not limited to the  model system.  

The influences of bosonic reservoirs 
	on the system are succinctly encoded
	in the spectral density
\begin{equation}
\gamma_v(\omega) = 2\pi \sum_k g^2_{k,v}\delta(\omega - \omega_{k,u}).
\end{equation}
In the continuum limit,	the spectral density can be assumed to take on the standard form
\begin{equation}	
\gamma_v(\omega) = \pi\alpha_v \omega^s \omega^{1-s}_{c,v} 
		\psi_c\left(\left.\omega\right/\omega_{c,v}\right)
\end{equation}
	with $\alpha_v$,
	the dimensionless system-reservoir coupling strength,
	and $\omega_{c,v}$,
	the cutoff frequency of the $v$--th bosonic reservoir.
In this work,
	we will focus on the Ohmic case,
	$s=1$,
	which features more prominent contributions from the low--frequency modes
	in comparison to the super-Ohmic models
	with $s=3$ analyzed in earlier works
	using NE-PTRE method.
We will also consider different cutoff functions
	$\psi_c\left(\left.\omega\right/\omega_{c,v}\right)$ 
	to illustrate the robustness of NE-VPTRE in providing consistently improved results
	on heat current calculations
	under a variety of model specifications.

A challenging question in  studying non-equilibrium transport is the definition of the effective temperature of the system. In the non-equilibrium set-up, the temperatures of the 
reservoirs
are well-defined, but the temperature
 of the system is not directly determined or even defined.  This posts a conceptual difficulty in extending variational polaron method  to
 a non-equlibrium setup.  To address this question,  the effective temperature is first introduced within the framework of the nonequilibrium steady state and then justified by means of a simple mapping procedure. 

To proceed, we make an ansatz that the NESS can be approximated by the zero-th order term (with respect to $H_I$) of Eq.~(\ref{eq:nessf})\cite{Hershfield.93.PRL},
\begin{eqnarray}\label{eq:ness}
 \rho_{NESS}   &=& 
\frac{\exp\left(-\beta_L H_B^L -\beta_R H_B^R - \bar\beta  H_s \right)}
	%----------------------------	
	{Tr[\exp\left(-\beta_L H_B^L -\beta_R H_B^R - \bar\beta  H_s \right)]},
\end{eqnarray}
where the effective temperature is given 
\begin{eqnarray}\label{eq:teff}
 T_{eff} = { (\alpha_L T_L + \alpha_R T_R) \over (\alpha_L+\alpha_R) }
\end{eqnarray}
where $\alpha_{\rm L}$ and $\alpha_{\rm R}$ are the coupling strengths 
		to the left and right baths,
		defined in Eq.~(6).
The choice of the effective temperature $T_{eff}$ is derived next in Sec.~\ref{sec:Teff}	
and also justified in the next paragraph.    
We note that Eq.~(\ref{eq:ness}) preserves the 
factorized form of the density matrix due to the absence of $H_{I}$. 
 At this point, we have turned the nonequilibrium  setup to an approximate but much more familiar equilibrium setting 
in order to derive variational polaron transformed master equation under the effective temperature 
$K_B T_{eff}= 1/\bar \beta$ 
with the canonical form of equilibrium density matrices. 

Alternatively , we can adopt a simple mapping procedure to determine the effective temperature. 
In the non-equilibrium setup illustrated in Fig.~1, 
the thermal effect of a reservoir on the system is characterized by the influence functional
\begin{eqnarray}
C_v(t)= { 2 \over \pi} \int \gamma_v(\omega) [\coth(\omega  T_v) \cos(t \omega) - i \sin(t \omega) ]d\omega
\end{eqnarray}
where $\gamma_v(\omega)$ is the spectral density.
We now  assume the two thermal 
reservoirs
couple to the system with the same functional form of spectral density 
but different coupling strengths and temperatures.   Then, the two 
reservoirs can be combined to a single  effective 
reservoir
 characterized by the total influence functional  $C_{eff}(t)=C_R(t)+C_L(t)$.
When the temperature difference is small ($\delta T/ \tilde{T} \ll 1$) and/or the temperatures are high 
($\beta h\omega \ll 1$),  we can approximately write 
\begin{eqnarray}
	C_R(t)+C_L(t) \approx   { 2\over \pi}  
			\int [\gamma_R(\omega) + \gamma_L(\omega)] 
				 [\coth(\omega  T_{eff}) \cos(\omega t) - i \sin(\omega t)] 
				 d\omega
\end{eqnarray}
where the effective temperature is given exactly as in Eq.~(\ref{eq:teff}).
Thus, we can map the non-equilibrium setup to an effective equilibrium setup such that the system relaxes to the equilibrium distribution characterized by the effective temperature. It is also possible to generalize the above discussion to a more general case where the two coupling  operators are not identical, 
i.e., non-communitive transport.

While we use the NESB model to illustrate our newly introduced approach to the nonequilibrium quantum transport, 
	we emphasize the essential assumption, 
	such as Eq.~(\ref{eq:ness}), is based on a rigorous theoretical 
	framework\cite{Hershfield.93.PRL} 
	and can be adapted to other related transport models.

For a general non-equilibrium transport set-up, the basic principle of non-equilibrium variational polaron transformation remains applicable, though the calculation can be more involved.  To go beyond the NESB model, we can consider a multi-site system, such as an exciton chain or a spin chain.\cite{Manzano.16.NJP} Then the effective temperature is not a constant, but becomes a function of the coordinate along the chain.  Often, we can find an analytical solution of the temperature profile in some limiting cases, such as the strong coupling limit and/or Markov limit. Then, we use the effective temperature profile to define the non-equilibrium Gibbs state and apply variational polaron evaluation. Though the choice of the reference system is flexible and can affect the accuracy of the theoretical prediction, the NE-VPTRE approach remains valid and general.

%%%%%%%%%%
\subsection{Non-equilibrium Fermi's Golden-Rule Rate and Effective Temperature}
\label{sec:Teff}
%%%%%%%%%
For the NESB model considered in this work, one can write down the exact Liouville equation\cite{Cao.00.JCP} for the
composite system projected in the system's local basis $\{\ket{1},\ket{2}\}$,
\begin{eqnarray}\label{eq:rho}
\dot{\rho}_{12}(t) &=& -i \mathcal{L}_{12}\rho_{12} + i V (\rho_1 - \rho_2), \nonumber \\
\dot{\rho}_{1}(t) &=& -i \mathcal{L}_{1}\rho_{1} + i V (\rho_{12} - \rho_{21}), \nonumber \\
\dot{\rho}_{2}(t) &=& -i \mathcal{L}_{2}\rho_{2} - i V (\rho_{12} - \rho_{21})
\end{eqnarray}
where the Liouville superoperators are defined as $\mathcal{L}_{1/2} A \equiv [H_{1/2},A]$,
$\mathcal{L}_{12}A = H_1 A - A H_2$ and $\rho_{12}=\rho_{21}^*$.  In this subsection, we note 
the notation changes: $V=\Delta/2$, $H_1 = 
\epsilon/2 + \sum_{v} H_1^v$ with $H_1^v \equiv H_B^v + \ket{1}\bra{1} B_v $ and
$H_2 \equiv \sum_{v} H_B^v - \ket{2}\bra{2} B_v $.

Assuming $\rho_{12}(0)=0$, one obtains a formal solution,
\begin{eqnarray}\label{eq:rho12}
\rho_{12}(t) & = & i \int^t_0 d\tau e^{-i\mathcal{L}_{12}\tau}.
\end{eqnarray}
Next we assume the density matrix elements remain factorized as $\rho_1(t) = P_1(t) \rho_{b,1}$ 
and $\rho_2(t) = P_2(t) \rho_{b,2}$ in which the two baths are separately equilibrated with the system
such that $\rho_{b,i} \propto \exp(-\beta_L H_i^L)\exp(-\beta_R H_i^R)$ with $i=1,2$.
Substituting Eq.~(\ref{eq:rho12}) into the last two equations in Eq.~(\ref{eq:rho}) and use the factorization
assumption to trace out the bath, we obtain a general Fermi golden rule rate equation
\begin{eqnarray}
\dot{P}_1(t) &=& -\int^t_0 K_+(t-t')P_1(t') dt'  \nonumber \\
& & + \int^t_0 K_-(t-t')P_2(t') dt'.
\end{eqnarray}
Imposing the Markov approximation and keep only the first order expansion of the rate kernel $K_{1/2}$,
one obtains the standard rate equation, 
\begin{eqnarray}
k_{\pm} & = & \int^\infty_0 d\tau K_{\pm}(\tau) \nonumber \\
& = & V^2 \int^\infty_0 d\tau e^{\mp i \epsilon \tau} \exp\left[-g_L(\tau)-g_R(\tau)\right], \nonumber \\
\end{eqnarray}
with the lineshape function given by
\begin{eqnarray}
g_v(t) & = & \frac{2}{\pi} \int d\omega \frac{\gamma_v(\omega)}{\omega^2}  \left[\left\{1-\cos(\omega t)\right\}\coth(\beta_v\omega/2)-i
\sin(\omega t) \right]. 
\end{eqnarray}
The above derivation follows closely the formulation of non-Markov quantum rate equation in Ref.~\cite{Cao.00.JCP}.

If one further takes a short time and high temperature expansion on all trigonometric functions
and $\coth(\beta_v\omega/2)$ for $g_v(t)$,
the rate constants can be obtained in generalized Marcus form\cite{Galen.16.PNAS} after performing a Gaussian integration.
\begin{eqnarray}
k_{\pm} = 2\pi V^2\sqrt{\frac{\bar\beta}{4\pi\alpha}}\exp\left(-\bar\beta\frac{(\alpha \pm \epsilon)^2}{4\alpha}\right),
\end{eqnarray}
with $T_{eff} = \frac{\alpha_L}{\alpha} T_L + \frac{\alpha_R}{\alpha} T_R$,
and $\bar\beta = 1/T_{eff}$.  
Since the rate constant is controlled by the effective temperature,
this motivates us to propose the ansatz that the system is characterized by the effective temperature
in Eq.~(\ref{eq:teff}).

%%%%
\subsection{The Nonequilibrium Variational Polaron Transform}
\label{variational}
%%%%

We now apply the variational polaron transformation to the NESB model.  
The generalized polaron displacement operator
is given by
\begin{eqnarray}\label{eq:ptsf}
U	& = & \exp\left[\sigma^z D \right] \nonumber
\\
	& = & \exp\left[\sigma^z \sum_{v,k}\frac{f_{k,v}}{\omega_{k,v}}
			\left(b^\dag_{k,v}-b_{k,v}\right)\right].
\end{eqnarray}
The transformed Hamiltonian 
	$\tilde H = U H U^\dag = \tilde H_s +\tilde H_{I}+ H_b $ with
\begin{eqnarray}
	\tilde H_s 
& = & \frac{\epsilon}{2}\sigma^z + \frac{\Delta_R}{2}\sigma^x 
	+ \sum_{k,v} \frac{f_{k,v}}{\omega_{k,v}} 
	\left(f_{k,v}- 2 g_{k,v}\right), 
\nonumber \\
	%- \frac{1}{\omega_{\nu k}}
	%(f^\chi_{\nu k} (g^\chi_{\nu k})^*+ (f^\chi_{\nu k})^* g^\chi_{\nu k} \right) \nonumber \\
	\tilde{H}_{I} 
& = & 
	\sum_{\alpha=x,y,z} V_\alpha \sigma^\alpha.
\end{eqnarray}
%%%%
The rotated system-bath interactions take on the form,
\begin{eqnarray}
	V_x &= &  \frac{\Delta^2}{2}\left(\cosh\left(2 D\right)-\eta\right), 
\nonumber \\
	V_y &=& i \frac{\Delta^2}{2}\sinh\left(2 D\right),
\nonumber \\
	V_z &=& \sum_{k,v}\left(g_{k,v}-f_{k,v}\right) (a^\dag_{k,v} + a_{k,v}),
%B_\chi & = & \sum_{\nu k} \left(\frac{f^\chi_{\nu k}}{\omega_{\nu k}}b^\dag_{\nu k} -\left(\frac{f^\chi_{\nu k}}{\omega_{\nu k}}\right)^*b_{\nu k} \right) 
\end{eqnarray}
where the displacement operator $D$ was defined in Eq.~(\ref{eq:ptsf}).
The renormalized tunneling matrix element $\Delta_R = \Delta \eta$ reflects the polaron dressing effects and 
$\eta$ is given in Eq.~(\ref{eq:eta}).

The displacement parameters, $\{f_{k,v}\}$, are determined by minimizing the ``effective'' free energy upper bound $A_B$ given by the Feynman-Bogoliubov inequality,
\begin{eqnarray}\label{eq:A}
	A \leq A_B = -\frac{1}{\beta} \ln \text{Tr}
	\left(e^{-\beta  \tilde H_s} \right)
	 + 
	\langle \tilde H_{I} \rangle_0 + \mathcal{O}(\langle \tilde H_{I}^2 \rangle_0),
\end{eqnarray}
	where the second term $\langle \tilde H_{I} \rangle_0 = 0$ 
	by construction and higher-order
	terms at the end of right hand side are ignored. 
If we write $f_{k,v} = g_{k,v}F_v(\omega_{k,v})$,
	then the minimization conditions, 
	$\partial A_B / \partial f_{k,v} = 0$, 
	leads to the set of self-consistent equations consisting of,
\begin{eqnarray}\label{eq:Fv}
	F_v(\omega) & = & 
	\left[1+\tanh
		\left(\frac{\bar\beta \Lambda}{2}\right)
		\coth
		\left(\frac{\beta_v\omega}{2}\right)
		\frac{(\Delta_R)^2}{\Lambda \omega} 
	\right]^{-1},
\end{eqnarray}
where $\Lambda = \sqrt{\epsilon^2 + \Delta_R^2}$ and the tunneling renormalization factor reads

\begin{eqnarray}\label{eq:eta}
	\eta = \exp
	\left(-\sum_{v} \int d\omega 
		\frac{\gamma_v(\omega)}{\pi\omega^2} F^2_v(\omega)
		\coth
		\left( \frac{\beta_v}{2} \omega \right) 
	\right).
\end{eqnarray}
While our discussion uses the nonequilibrium spin-boson (NESB) model\cite{Leggett.87.RMP,Weiss.12.NULL} for illustration, we emphasize Eqs.~(\ref{eq:ness})-(\ref{eq:Fv}) can be generalized for other models. 

Following the derivation\cite{Breuer.07.NULL} of the Born-Markovian Redfield equation 
in the polaron picture\cite{McCutcheon.11.PRB,McCutcheon.11.JCP,Lee.15.JCP,Wang.15.SR,Xu.16.FP}, %,Sun.16.JCP 
one obtains 
\begin{eqnarray}\label{eq:ptre}
	&& \frac{d \rho_s}{d t}=-i 
		\left[H_s, \rho_s \right]+  
\nonumber \\
	&& \,\, 
		\sum_{\substack{\alpha,\beta= \\ \{x,y,z\}}}
		\sum_{\substack{\omega,\omega^\prime= \\  0,\pm\Lambda}}
		\left(
			 \Gamma_{\alpha\beta,+}(\omega)
			+\Gamma_{\alpha\beta,-}(\omega^\prime)		
		\right) 
		P_{\beta}(\omega)\rho_s 
		P_{\alpha}(\omega^\prime)
	\notag\\
	&& \quad
		-\sum_{\substack{\alpha,\beta= \\ \{x,y,z\}}}
		 \sum_{\substack{\omega,\omega^\prime= \\ 0,\pm\Lambda}} 
		\left(\Gamma_{\alpha\beta,+}(\omega)P_\alpha(\omega^\prime)
				P_\beta(\omega)\rho_s + h.c. 
		\right),
%\Gamma_{\alpha\beta,-}(\omega) \rho_s P_\beta(\omega) P_\alpha(\omega^\prime) \right),
\end{eqnarray}
where the transition rates $\Gamma_{\alpha\beta,\pm}(\omega)$ % \equiv \int^\infty_0 d\tau e^{i\omega\tau} \langle V^\alpha(\pm\tau)V^\beta (0)\rangle_0$ 
are provided explicitly in Supplementary Materials and the eigen-basis decomposition of Pauli matrices in the interaction picture gives 
$\sigma^\alpha (-\tau)=\sum_{\omega=0,\pm\Lambda} 
	P_{\alpha}(\omega)e^{i\omega\tau}$.

It has been observed that the variational polaron transformation may suffer from sharp changes in the renormalized tunneling constant,%
\cite{Silbey.84.JCP, McCutcheon.11.PRB, Lee.12.JCP} which can lead to difficulties in the prediction of density matrix propagation. However, the numerical results reported in the next section suggest that the VPTRE prediction of steady-state heat current does not suffer from the discontinuity problem. In essence, heat current a non-equilibrium steady-state (NESS) solution and is thus more related to thermal equilibrium in the long-time limit than to dynamic coherence at finite times. In this context, the thermodynamic consistency of PTRE or VPTRE 
established early\cite{Xu.16.FP} can help explain the accuracy of the heat current prediction reported next.

%%%%%%%%%%%%%%%%%%%%%
\subsection{Heat Current and Full Counting Statistics}
%%%%%%%%%%%%%%%%%%%%%
	The definition of heat current is formalized through a two-time measurement 
protocol\cite{Esposito.09.RMP,Campisi.11.RMP}: 
	at time $t=0$, an initial measurement implemented via a projector 
	$K_{q_0}=|q_0\rangle\langle q_0|$ 
	to determine the energy content of 
	$H_B^R=\sum_{k}\omega_{k,R}b^{\dagger}_{k,R}b_{k,R}$ 
	and get an outcome $q_0$. 
A second measurement is performed at a later time $t>0$ 
	with another projector $K_{q_t}=|q_t\rangle\langle q_t|$ that 
	gives an outcome $q_t$. 
Hence, the net transferred heat is determined by $Q(t)=q_t-q_0$. 
The joint probability to measure $q_0$ and $q_t$ at two specified time points reads 
\begin{equation}
	P[q_t,q_0]~\equiv~\mathrm{Tr}
	\{K_{q_t} U(t,0) K_{q_0}\rho(0) K_{q_0} U^{\dagger}(t,0) K_{q_t}\},
\end{equation}
	where $U(t,0)$ the unitary time evolution operator of the total system 
	and $\rho(0)$ is the initial density matrix. 
One can further define the probability distribution for the net transferred quantity, 
	$Q(t)=q_t-q_0$ over a period of time $t$, 
\begin{equation}\label{eq:pqt}
	p(Q,t) = \sum_{q_t,q_0}\delta(Q(t)-(q_t-q_0))P[q_t,q_0],
\end{equation}
where $\delta(x)$ denotes the Dirac distribution. The corresponding cumulant generating function (CGF) for 
$p(Q,t)$ reads
%%%
\begin{equation}\label{eq:cgf}
G(\chi,t)~=~\ln \int\,dQ(t) p(Q,t)e^{i\chi Q(t)}.
\end{equation}
with $\chi$ the counting-field parameter. 
From CGF $G(\chi,t)$, one obtains an arbitrary $n$-th order cumulant of heat transfer via 
\begin{eqnarray}
	J^{(n)}(t) =
	\left.
		\frac{\partial^n G(\chi,t)}
			%-----------------------
			{\partial(i\chi)^n}
	\right\vert_{\chi=0}
\end{eqnarray}
Notably, the first and second cumulants correspond to the heat current and its noise power.

If the reduced density matrix (RDM) $\rho_s$ 
	is decomposed into different subspaces of net transferred $Q$, i.e. 
	$\rho_s(t) = \int dQ \rho_s(Q,t)$, 
	then $p(Q,t) = \text{Tr}_s \rho_s(Q,t)$.
Substituting the $Q$-resolved RDM into Eq.~\ref{eq:cgf}, 
	one may further define $\chi$-resolved RDM,
\begin{eqnarray}
\rho_s^\chi(t) = \int dQ e^{i\chi Q} \rho_s(Q,t).
\end{eqnarray} 
The statistics of heat transfer can be extracted from $\rho_s(\chi,t)$ via the relation 
$G(\chi,t) = \ln \text{Tr}_s \rho_s^\chi(t)$.

The dynamics of $\rho^\chi_s(t)$ 
	is obtained after a counting field $\chi$ is introduced via a unitary transformation
	$U_\chi = \exp(-iH_b^R\chi/2)$ 
	to count the net amount of transferred energy into the right bath.  
Without delving into further derivations, 
	which is a straightforward generalization of earlier works,
	we get a $\chi$-dependent NE-VPTRE,
\begin{eqnarray}\label{eq:vptre}
	& & \frac{d \rho_s^\chi}{d t} 
		= -i \left[H_s, \rho_s^\chi \right]+ 
\nonumber \\
	& & \,\, \sum_{\substack{j,k= \\ \{x,y,z\}}}
			 \sum_{\substack{\omega,\omega^\prime= \\ 0,\pm\Lambda}}
		\left(
			\Gamma_{jk,+}^\chi(\omega) + \Gamma_{jk,-}^\chi(\omega^\prime)
		\right) 
		P_{k}(\omega)\rho_s^\chi P_{j} (\omega^\prime)
\notag\\
	&&\quad -\sum_{\substack{j,k= \\ \{x,y,z\}}}
			 \sum_{\substack{\omega,\omega^\prime= \\ 0,\pm\Lambda}} 
		\left(
			\Gamma_{jk,+}(\omega)P_j(\omega^\prime)
			P_k(\omega)\rho_s^\chi + h.c. 
		\right),
%kim --> definition of h.c.
%+ \Gamma_{jk,-}(\omega) \rho_s P_k(\omega) P_j(\omega^\prime) 
%& & = \,\, \mathcal{L}_\chi \rho_s
%\Gamma^{\chi}_{kj,+}\rho_s P_k(-\omega)P_j(-\omega^\prime)
\end{eqnarray}

By adopting the Liouville-space notation, Eq.~(\ref{eq:vptre}) could be succinctly 
expressed as $d\rho_s^\chi(t)/dt = -\mathcal{L}_\chi \rho_s^\chi(t)$.
The formal solution assumes the simple form $\rho_s^\chi(t) = e^{-\mathcal{L}_\chi t} \rho_s^\chi(0)$ when
the Hamiltonian is time-independent.  
The stationary CGF at the steady state is obtained via $G(\chi) = \lim_{t\rightarrow\infty}\frac{1}{t}\ln \text{Tr}_s \rho_s^\chi(t)$.  In the asymptotic limit\cite{Wang.14.NJP,Wang.15.SR}, one can show that $G(\chi) = -E_0(\chi)$ where $E_0(\chi)$ is the ground state of the superoperator $\mathcal{L}_\chi$.  Hence, the heat current
can be conveniently calculated via
\begin{eqnarray}
	J_R & = & 
	-\left.	\frac{\partial E_0(\chi)}
			     {\partial (i\chi)}
	\right\vert_{\chi=0}.
\end{eqnarray}

In the case of the unbiased NESB model $\epsilon=0$, 
	we derive the following expression for the heat current
%%
%\begin{widetext}
\begin{eqnarray}\label{eq:unbiasedJ}
J & = & \frac{1}{2\pi} \int d\omega \, \omega \left[C_{xx}(0,\omega) + (C_{yy}(\Lambda,\omega)+C_{zz}(\Lambda,\omega)) \Phi^+ 
(C_{yy}(-\Lambda,\omega)
\right.
\notag\\
&\quad&  \left. +C_{zz}(-\Lambda,\omega))\Phi^- \right],
\end{eqnarray}
%\end{widetext}
%
where the renormalized energy gap defined earlier reduces to 
	$\Lambda=\Delta\eta$ for $\epsilon=0$, and
\begin{eqnarray}
\Phi^{\pm} = \frac{\Sigma_d^2+\Delta_{od}^2\pm\left(\Sigma_d\Delta_d+\Sigma_{od}\Delta_{od}\right)}{\Sigma_d^2+\Delta_{od}^2},
\end{eqnarray}
with 
	$\Sigma_d = \sum_{\nu=y,z}\phi_{\nu\nu}(\Lambda)+\phi_{\nu\nu}(-\Lambda)$,
	$\Sigma_{od} = \sum_{\nu=y,z}\phi_{\nu\bar \nu}(\Lambda)+\phi_{\nu\bar \nu}
		(-\Lambda)$,
	$\Delta_d = \sum_{\nu=y,z}\phi_{\nu\nu}(\Lambda)-\phi_{\nu\nu}(-\Lambda)$ 
and
	$\Delta_{od}= \sum_{\nu=y,z}\phi_{\nu\bar \nu}(\Lambda)-\phi_{\nu\bar \nu}
		(-\Lambda)$.
Note the subscript satisfies $\bar{\nu}=z / y$ if $\nu=y / z$.
The correlation functions read,
\begin{eqnarray}
\phi_{\nu\mu}(\omega) & = & \int dt e^{\i\omega t} C_{\nu\mu}(t) \nonumber \\
& = & \int d \omega  C_{\nu\mu}(\omega,\omega'),
\end{eqnarray}
where $C_{\nu\mu}(t)$ are explicitly given in the supplementary information, and $C_{\nu\mu}(\omega,\omega')$ are defined on the second line of this equation.

In the weak and strong coupling regimes, Eq.~(\ref{eq:unbiasedJ}) reduces smoothly back to the Redfield and NIBA
results, respectively. In the weak coupling limit, we note the polaron displacement $F_v(\omega) \rightarrow 0 $,which subsequently leads to the correlation functions $C_{xx}(0,\omega)\rightarrow 0$ and $C_{yy}(\pm\Lambda,\omega)\rightarrow 0$. In this case, only the two terms, $C_{zz}(\pm \Lambda,\omega)$, 
survive and approach the Redfield result. 
On the other hand, in the strong coupling limit, $F_v(\omega) \rightarrow 1$ yield a full polaron displacement while
the last two terms of Eq.~(\ref{eq:unbiasedJ}) vanish. In this case, Eq.~(\ref{eq:unbiasedJ}) reduces to NIBA results in the strong coupling limit.  In the next section, we will numerically demonstrate the limiting behaviors of
Eq.~(\ref{eq:unbiasedJ}) in various examples.

%:Figure 2
%%
\begin{center}
\begin{figure}
\includegraphics[width=1\columnwidth]{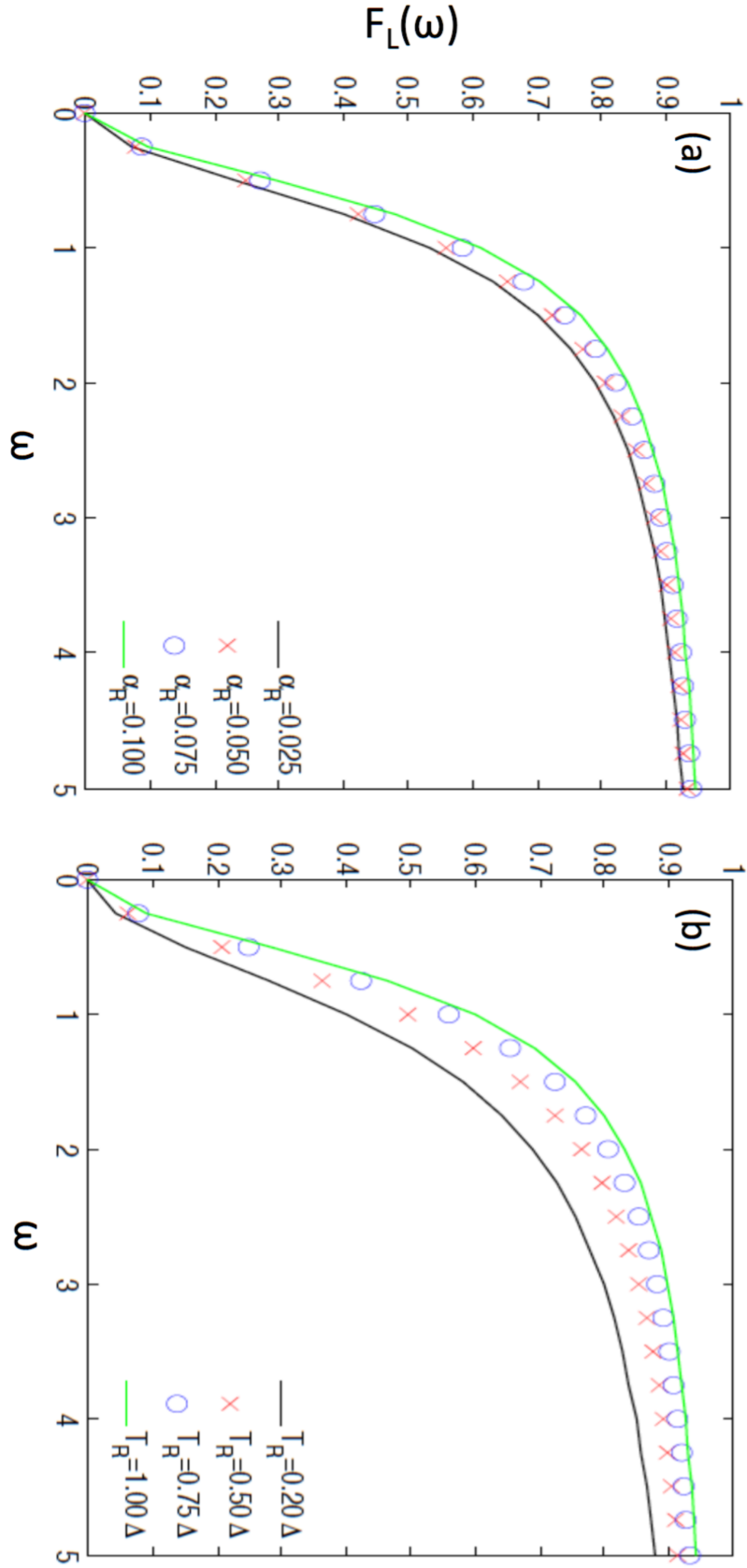}
\caption{Cross-bath correlation effects on $F_L(\omega)$, Eq.~(\ref{eq:Fv}), for an unbiased NESB model with parameters: $\omega_c=10\Delta$, $\alpha_L=0.05$ and $T_L=\Delta$. (a): variation of $\alpha_R$ while $T_R=0.75 \Delta$ is fixed.  (b): variation of $T_R$ while $\alpha_R=0.05$ is held fixed.}
\label{fig:Fv}
\end{figure}
\end{center}

%:Figure 3
%%
\begin{center}
\begin{figure}
\includegraphics[width=0.75\columnwidth]{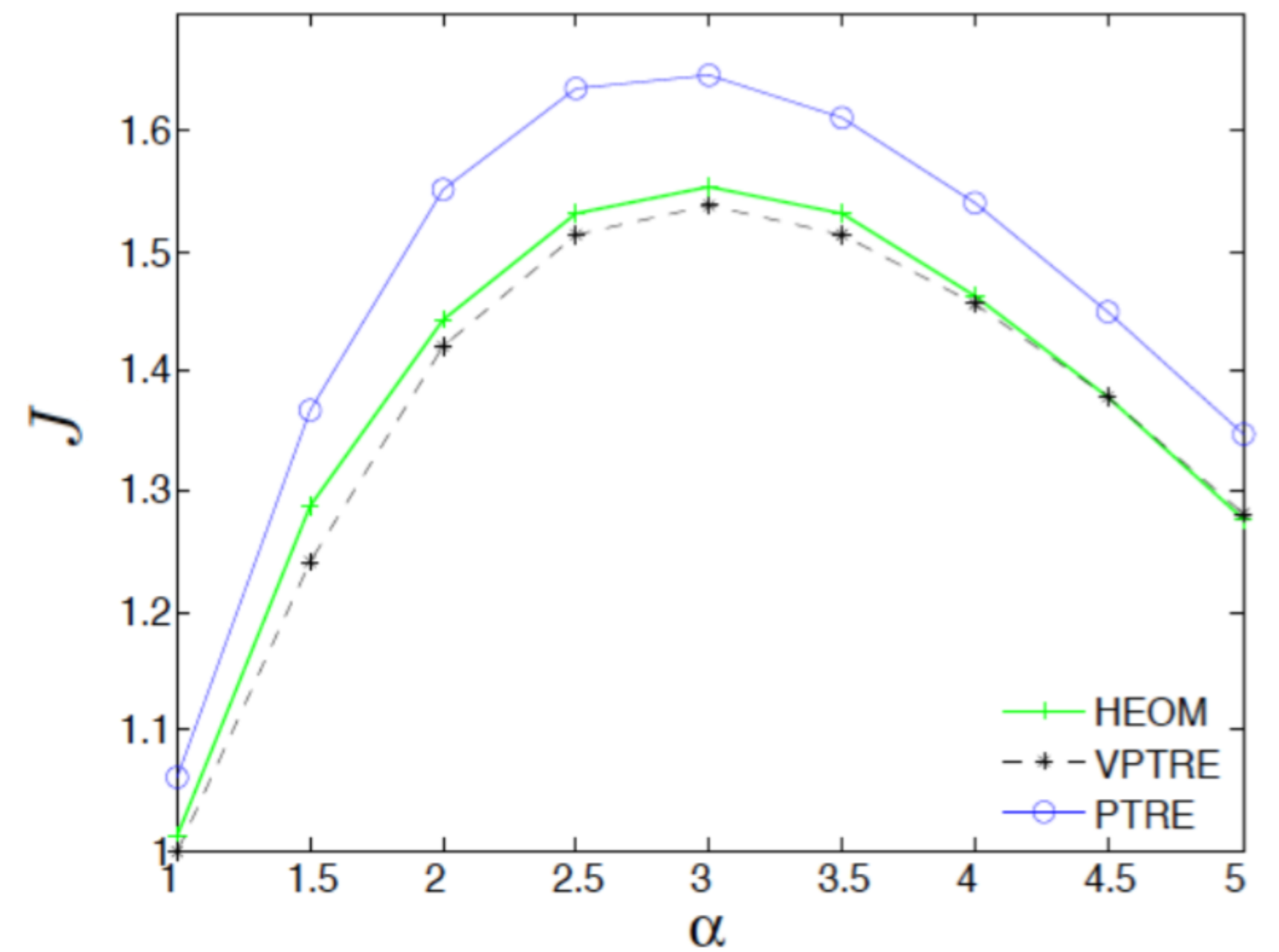}
\caption{The heat current as a function of coupling strength.  The NESB model parameters
are $\Delta/\omega_c=1/16$,$\epsilon/\omega_c=1/4$, $T_L/\omega_c=0.275$, $T_R/\omega_c=0.225$t.
The spectral density is super-ohmic with s=3 and a rational cutoff. }
\label{fig:superOhmic}
\end{figure}
\end{center}
%%

%%
%%%%%%%%%%%%%%
\section{Results}\label{sec:results}
%%%%%%%%%%%%%%
%As discussed in the introduction, the NE-PTRE method is a powerful approach to analyze statistics of heat transfer through low-dimensional quantum systems coupled to high-temperature and fast-relaxing heat reservoirs. In the super-Ohmic cases, the full polaron transformation provides a unified theory for heat transfer.  In this section, we consider Ohmic bath models and illustrate how the variationally displaced polarons improve the polaron-based heat current calculations beyond super-Ohmic bath models.

%%%%%%%%%%%%%
\subsection{Nonequilibrium Variational Polaron Displacement: Cross-bath Correlation}
%%%%%%%%%%%%%%
We first take a closer look at the nonequilibrium variational polaron displacements proposed in the Method section.
The functional form of $F_v(\omega)$ in Eq.~(\ref{eq:Fv}) is almost identical to the standard equilibrium forms.
Hence, many of the equilibrium results carry over to the nonequilibrium set-ups.  For instance, $F_v(\omega)$ prescribes small displacements, $f_{vk} \approx 0$, for slow modes and close to the full polaron displacements, $f_{vk}=g_{vk}$, for fast modes. Due to additivity nature of the exponents representing contributions from different baths in Eq.~(\ref{eq:eta}), the renormalized tunneling factor $\eta$ can be approximately factorized into two components to reflect interaction with two baths, respectively.  Because of this resembleance to equilibrium results, we can clearly see that the original NE-PTRE works well for high-temperature and fast-relaxing heat reservoirs when $\eta \ll 1$.

The major distinction that sets apart nonequilibrium variational method is the cross-bath correlation effects mediated by the central spin via the hyperbolic tangent factor, $\tanh\left(\frac{\bar\beta \Lambda}{2}\right)$, evaluated at the weighted average of the inverse temperature $\bar\beta$ in Eq.~(\ref{eq:Fv}).  In Fig.~\ref{fig:Fv}, the displacements of the low-frequency modes in the left bath clearly depend on the parameters of the right bath as long as $\Lambda \neq 0$, i.e. the two system eigenstates are not degenerate.  When the right bath is more strongly coupled to the system or is elevated to a higher temperature, the displacements of left bath modes also become more pronounced as displayed in the figure.

We first consider an NESB model with two super-Ohmic baths characterized by a rational cutoff function 
$\psi_c(\omega/\omega_{c,v}) = \frac{1}{(1+(\omega^2/\omega_c^2))^2}$. 
As demonstrated in Fig.~\ref{fig:superOhmic}, the new NE-VPTRE results are in excellent agreement with the numerically exact hierarchical equation of motion\cite{tanimura.06.jpsj,Kato.16.JCP,hsieh.18.jcp,duan.17.prb} (HEOM) data than the original NE-PTRE method. In the intermediate coupling strength regime presented in Fig.~\ref{fig:superOhmic}, the full polaron displacement overestimates the heat current.  
This discrepancy is precisely due to the inaccurate treatment of the low-frequency modes in the baths.
This observation can be confirmed by increasing the cutoff frequency $\omega_{c,v}$ of the bath, the dissipations
induced by the low-frequency modes are diluted and the variational
 functions $F_{L/R}(\omega)$, Eq.~(\ref{eq:Fv}), approach a constant unity.  The results of the two polaron methods then converge in this limit.

%:Figure 4
%%
\begin{center}
\begin{figure}
	\includegraphics[width=1\columnwidth,trim=4 4 4 4,clip]{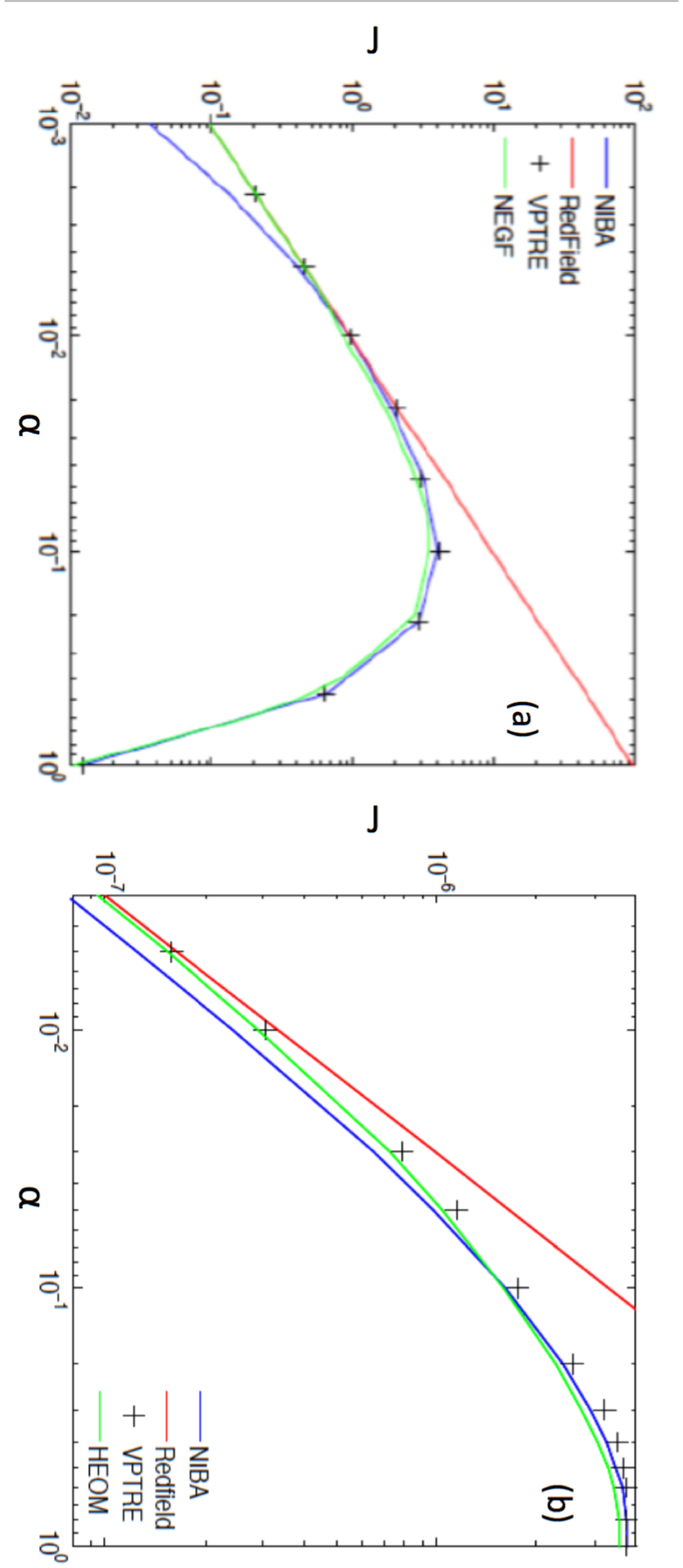}
%[]%width=0.9\columnwidth
\caption{The heat current as a function of coupling strength. (a) model parameters: $\Delta=\omega_c/30$, $T_L=1.4\Delta$, $T_R=1.2\Delta$ and $\epsilon=0$. (b) model parameters: $\Delta=0.02\omega_c$, $T_L=1.0\omega_c$, $T_R=0.9\omega_c$ and $\epsilon=0$. For the Ohmic bath models, NIBA and NE-PTRE are equivalent.}
\label{fig:ohmic}
\end{figure}
\end{center}

%%

%:Figure 5
%%
\begin{center}
\begin{figure}
	\includegraphics[width=1\columnwidth,trim=4 4 4 4,clip]{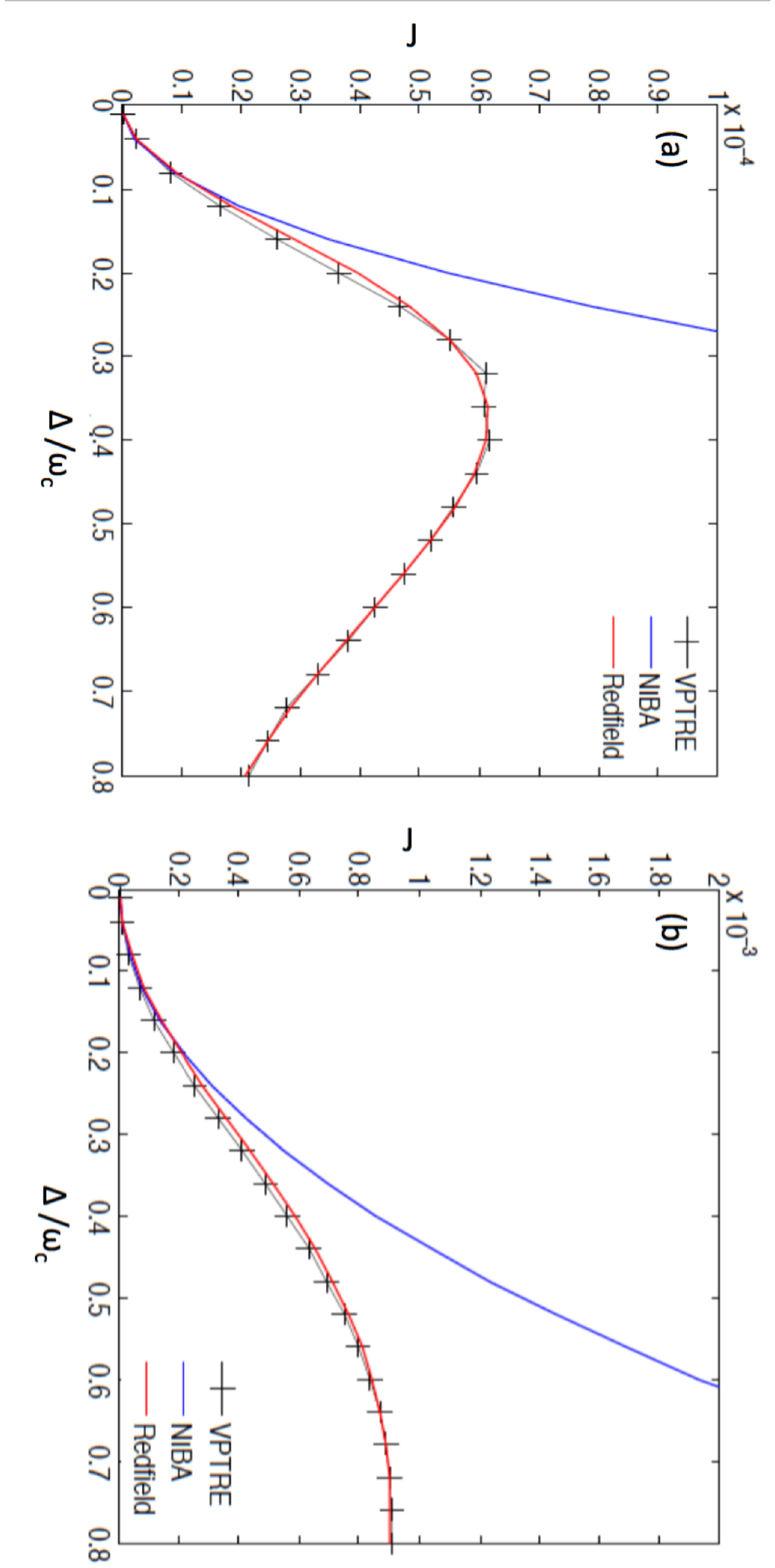}
\caption{The heat current as a function of $\Delta$ with fixed $\omega_c$. 
(a) Model parameters: $\alpha=0.03$, $T_L=0.15 \omega_c$, $T_R=0.14 \omega_c$.
(b) Model parameters: $\alpha=0.03$, $T_L=0.4 \omega_c$, $T_R=0.3 \omega_c$.
For the Ohmic bath models, NIBA and NE-PTRE are equivalent.}
\label{fig:time}
\end{figure}
\end{center}

%%

%%%%%%%%%%%%%
\subsection{Unified Heat Current Calculation for Ohmic Spectral Density}
%%%%%%%%%%%%%%
Next, we turn to the NESB cases in which heat reservoirs are featured with Ohmic spectral densities.
The NE-PTRE reduces exactly to NIBA regardless of the
cutoff function $\psi_{c}(\omega/\omega_{c,v})$. Hence, the full polaron displacement can only capture the incoherent part of the heat current and provide an accurate results only in the strong coupling and/or scaling limits.

The variational method extends the applicability of the polaron method beyond super-Ohmic cases.
To demonstrate the improvement, we first calculate the heat currents as function of the coupling strength in Fig.~\ref{fig:ohmic}. Different spectral cutoff functions are used in the panel a: $\psi_v(\omega)=\frac{1}{(1+(\omega^2/\omega_c^2))^4}$ and  panel b: $\psi_v(\omega)=e^{-\omega/\omega_c}$, respectively. Over the entire range
coupling strength considered, the NE-VPTRE agrees well with the exact result (i.e. NEGF in panel a and HEOM in panel b). In the weak-coupling limit, the exact results approach the Redfield result; while the NE-PTRE method (equivalent to NIBA for Ohmic baths) do not fare well. These two results establish the superiority of NE-VPTRE over the NE-PTRE\cite{Wang.15.SR} in dealing with Ohmic baths.

We next investigate the heat current as a function of the spin tunneling frequency 
	$\Delta$ in Fig.~\ref{fig:time}. 
The two baths are taken to be an identical Ohmic form,
	characterized by an exponential cutoff function, 
	but held at different temperatures.  
A weak system-bath coupling strength is used for the two cases presented in Fig.~\ref{fig:time} such that 
the Redfiled results provide reasonably accurate benchmarks within the specified range of $\Delta/\omega_c$. 
Under both small (panel a) and large (panel b) thermal bias, an excellent agreement between Redfield and NE-VPTRE affirm the applicability of the variational polaron-based framework to calculate heat current beyond the scaling limit, i.e. $\Delta/\omega_c \ll 1$. The essential need of adopting a variational polaron displacement is further supported by the observation on how the NE-PTRE (or NIBA) fails to capture the turnover behaviors portrayed in 
Fig.~\ref{fig:time} because it only accounts for the spin tunneling up to $\Delta^2$ in the heat current calculation. On the other hand, the variational polaron result takes into account of higher order terms of $\Delta$
in the weak system-bath coupling limit. 

The turnover can be more transparently explained via the Redfield expression for heat current,
\begin{eqnarray}\label{eq:redfield}
J & = & \frac{\pi\Delta}{8} \frac{\gamma_L(\Delta)\gamma_R(\Delta)}{\gamma_L(\Delta)(n_L(\Delta)+1/2)+\gamma_R(\Delta)(n_R(\Delta)+1/2)} \nonumber \\
& & \times \left(n_R(\Delta)-n_L(\Delta)\right),
\end{eqnarray}
where $n_v(\omega)$ is the Bose-Einstein distribution. In the Redfield framework, a classical-like sequential energy transfer is at work, and only bath modes in resonance (i.e. $\omega = \Delta$) with the system contribute to the heat conduction. Since we only consider parameter regime $\Delta < \omega_c$ for both cases considered in Fig.~\ref{fig:time}, the Ohmic spectral densities $\gamma_v(\Delta)$ are monotonically increasing with respect to $\Delta$. The turnovers of the heat current in Fig.~\ref{fig:time} are entirely controlled by the Bose-Einstein distributions in Eq.~(\ref{eq:redfield}). In short, the current diminishes once $\Delta$ exceeds the thermal energy appreciably such that the resonant modes in the bath are not thermally excited. The shift of the peak current to higher value of $\Delta$ (for the higher temperature case in Fig.~\ref{fig:time}b) also supports the claim that the heat conduction depends largely on the temperatures in this case.

%:Figure 6
%%
\begin{center}
\begin{figure}
\includegraphics[width=0.75\columnwidth]{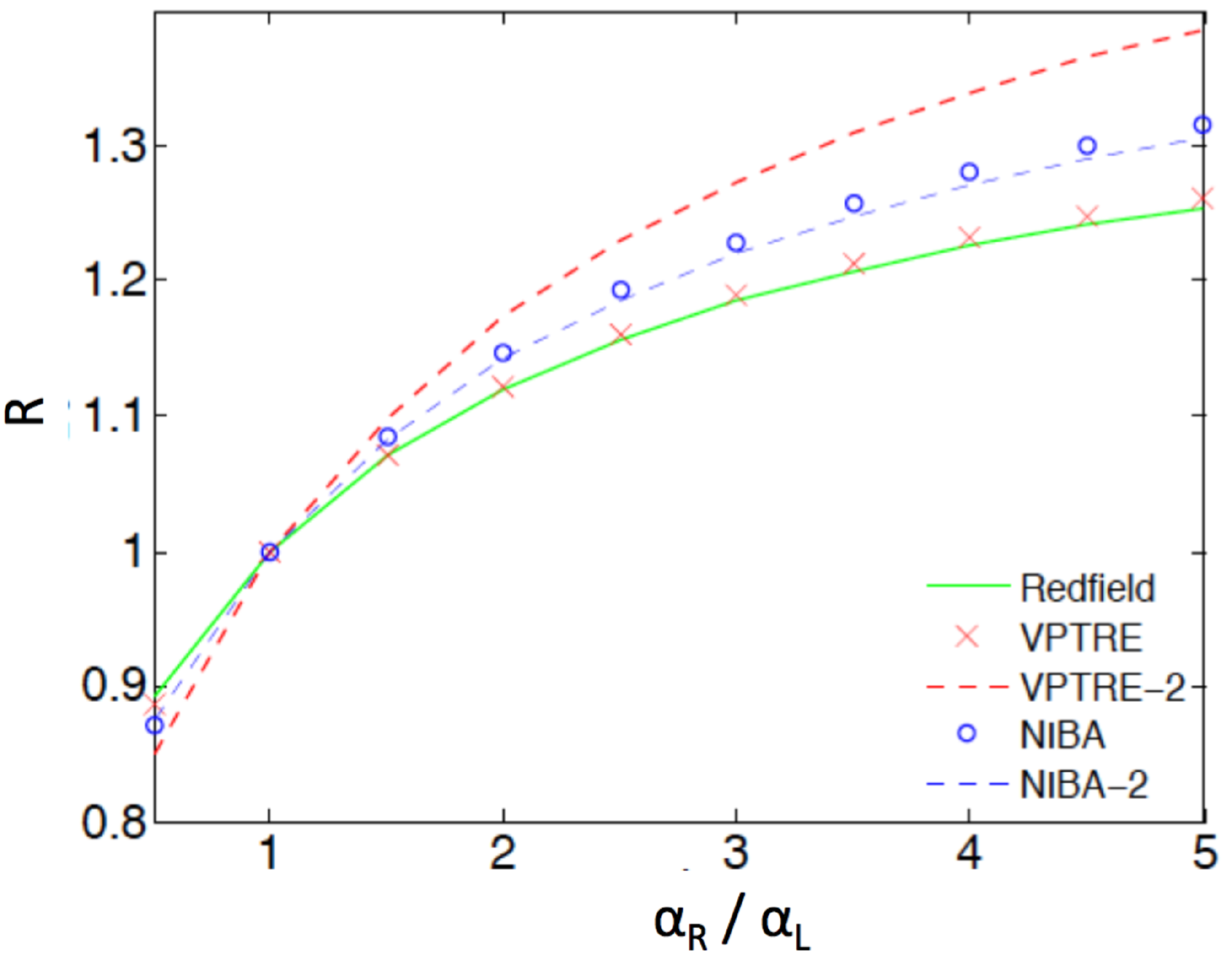}
\caption{Thermal rectification ratio as a function of $\alpha_R/\alpha_L$ asymmetry.
	The model parameters are $\omega_c=10\Delta$, $\epsilon = 0$, $\delta T=\vert T_L-T_R \vert = 0.4\Delta$ 
	and $T_{avg}=(T_L+T_R)/2 =\Delta$.
	The ``red cross" and red dashed lines are VPTRE results for case (1): $\alpha=\alpha_L+\alpha_R=0.05$ 
	and case (2): $\alpha=\alpha_L+\alpha_R=0.20$, respectively. The ``blue cross" and blue dashed lines are
	NIBA results for the same case (1) and case (2), respectively. For Ohmic bath models, NIBA and NE-PTRE
	are identical.}
\label{fig:rect}	
\end{figure}
\end{center}
%%

%%%%%%%%%%%%%
\subsection{Thermal Rectifications}
%%%%%%%%%%%%%%

In this section, we consider an NESB model where the two Ohmic baths have
	asymmetrical coupling strengths, 
	i.e. $\alpha_L \neq \alpha_R$.  
Because the spin is an anharmonic system, 
	a rectification can arise when the temperature bias $T_L-T_R=\delta T$ 
	on the two asymmetric baths are switched to $T_L-T_R=-\delta T$.  
More precisely, we define the rectification ratio  $R \equiv \left\vert J_R(\delta T) /  J_R(-\delta T) \right\vert$ under the constraint of a fixed $\alpha=\alpha_L + \alpha_R$ to remove the dependence of the rectification ratio $R$ on the overall magnitude of the coupling strengths, $\alpha$. The rectification ratio is a useful indicator to refelect whether the central quantum system is an ideal thermal diode\cite{Segal.05.PRL,Segal.16.ARPC}.

In Fig. \ref{fig:rect}, we consider two  overall magnitudes of the coupling strengths: 
$\alpha=0.05$ and $\alpha=0.2$.  For both cases, the same Redfield result (green curve in Fig.~\ref{fig:rect}) is obtained. This seemingly universal rectification behavior is an artifact of approximations and can be seen from Eq.~(\ref{eq:redfield}). Given the linearity of the current with respect to $\alpha$ in Eq.~(\ref{eq:redfield}), the rectification ratio R only depends on the ratio of $\alpha_R/\alpha_L$ and not their overall magnitude $\alpha$.
Nevertheless, the Redfield result is reliable for the weak value case such as $\alpha=0.05$. Indeed, the NE-VPTRE result (red cross) also agrees well with the Redfield.  For larger coupling case ($\alpha=0.2$),
the NE-VPTRE predicts an enhanced rectification ratio as typically expected for an anharmonic junction.
On the other hand, the two NIBA results 
	(blue circle and dashed line corresponding to $\alpha=0.05$ and $0.20$,
		respectively) 
	also collapse onto the same line in Fig.~\ref{fig:rect}.  
This is mainly due to the set of parameters chosen for illustration. 
In particular, $\omega_c=10\Delta$ is at the borderline of the scaling limit.  
When applying NIBA outside their valid parameter regimes, 
	the deviation from accurate results could be significant 
	as manifested by the rectification calculation shown in Fig.~\ref{fig:rect} 
	as well as Fig.~\ref{fig:time}. 

In this case, the superiority of NE-VPTRE is attributed to its better handling of the low-frequency modes through reduced polaron displacements.  When $\omega_c$ is further increased, the overall contributions from the low-frequency modes is diluted, and one would find NIBA results on $R$ to agree better with that of NE-VPTRE.

%%%%%%%%%%%%%
\section{Discussion}\label{sec:con}
%%%%%%%%%%%%%%
In summary, we extend the nonequilibrium polaron transformed Redfield equation (NE-PTRE) framework by variationally tuning the polaron displacements.
The generalization of a free-energy-based variational principle and Feynman-Bogoliubov inequality is built upon
the ansatz, Eq.~(\ref{eq:ness}), that the zero-th order nonequilibrium steady state of the composite system assumes a form resembling an equilibrium density matrix.  Similar to the equilibrium cases, the variational method extends the usefulness of a polaron picture beyond the nonadiabatic limit for nonequilibrium processes.
This achievement allows us to formulate a heat transfer theory beyond the super-Ohmic models in the polaron picture. 

In the Result section, we explicitly demonstrate several improvements of the newly proposed NE-VPTRE method in the aforementioned circumstances. Specifically, we observe that (1) improved numerical accuracy for super-Ohmic bath models over a broader range of model parameters. (2) Correctly recover the coherent tunneling effects on the heat current for the Ohmic bath models beyond the scaling limit as illustrated in Fig.~\ref{fig:time}.  (3) For calculations of rectification, an important indicator for the quality of a thermal diode, NE-VPTRE provides absolute improvements over both Redfield and the original NE-PTRE (equivalent to NIBA) methods in the Ohmic cases. 

The present work not only significantly extends the applicability of a transparent and numerically efficient polaron picture to calculate heat current but also builds a foundation to calculate higher order statistics\cite{Wang.17.PRA,Liu.17.JCP} of heat transfer, such as the noise power spectrum relating to the fluctuations of heat transfer. Future work will generalize the present framework to handle non-Markovian effects 
%\cite{Braggio.06.PRL,Flindt.08.PRL,Flindt.10.PRB} 
and to formulate a unified energy transfer theory incorporating the low-temperature regime and to investigate fluctuations.

\section*{Acknowledgements}

All authors  acknowledge the support from Singapore-MIT Alliance for Research and Technology (SMART). 
J. C. acknowledges the support from the National Science Foundation (NSF) (Grant CHE 1836913 and CHE 1800301).%

%%%%%%%%%%%%%%%%%%%%%%
%\bibliographystyle{apsrev} 
%\bibliography{heat}

%\newpage
%\listoffigures

\setcounter{equation}{0}
\renewcommand{\theequation}{B\arabic{equation}}

\end{document}